\documentclass[%
reprint,
showkeys,
amsmath,amssymb,
aps, physrev,
prl,
]{revtex4-2}

\usepackage[T1]{fontenc}
\usepackage[utf8]{inputenc}
\usepackage[main=french]{babel}
\usepackage[colorlinks = true,
            linkcolor = blue,
            urlcolor  = blue,
            citecolor =  blue,
            anchorcolor = blue]{hyperref}
\usepackage{graphicx,float}
\usepackage{xcolor}

\usepackage{bm}
\usepackage{dsfont}
\usepackage{stmaryrd}
\usepackage{tikz}
\usepackage{mdframed}
\usepackage[most]{tcolorbox}
\tcbuselibrary{breakable}
\usepackage{lipsum}
\usepackage[ruled,vlined,linesnumbered]{algorithm2e}
\usepackage{algpseudocode}
\usepackage{pgfplots}

\newcommand{\ds}{\mathrm{d}}
\newcommand{\p}{\partial}
\newcommand{\bb}{\mathbb}
\newcommand{\ff}{\mathfrak}
\newcommand{\cc}{\mathcal}
\RequirePackage[normalem]{ulem} 
\RequirePackage{color}\definecolor{RED}{rgb}{1,0,0}\definecolor{BLUE}{rgb}{0,0,1} 
\providecommand{\DIFaddbegin}{} 
\providecommand{\DIFaddend}{} 
\providecommand{\DIFdelbegin}{} 
\providecommand{\DIFdelend}{} 
\providecommand{\DIFaddbeginFL}{} 
\providecommand{\DIFaddendFL}{} 
\providecommand{\DIFdelbeginFL}{} 
\providecommand{\DIFdelendFL}{} 
\newcommand{\DIFscaledelfig}{0.5}
\RequirePackage{settobox} 
\RequirePackage{letltxmacro} 
\newsavebox{\DIFdelgraphicsbox} 
\newlength{\DIFdelgraphicswidth} 
\newlength{\DIFdelgraphicsheight} 
\LetLtxMacro{\DIFOincludegraphics}{\includegraphics} 
\newcommand{\DIFaddincludegraphics}[2][]{{\color{blue}\fbox{\DIFOincludegraphics[#1]{#2}}}} 
\newcommand{\DIFdelincludegraphics}[2][]{
\sbox{\DIFdelgraphicsbox}{\DIFOincludegraphics[#1]{#2}}
\settoboxwidth{\DIFdelgraphicswidth}{\DIFdelgraphicsbox} 
\settoboxtotalheight{\DIFdelgraphicsheight}{\DIFdelgraphicsbox} 
\scalebox{\DIFscaledelfig}{
\parbox[b]{\DIFdelgraphicswidth}{\usebox{\DIFdelgraphicsbox}\\[-\baselineskip] \rule{\DIFdelgraphicswidth}{0em}}\llap{\resizebox{\DIFdelgraphicswidth}{\DIFdelgraphicsheight}{
\setlength{\unitlength}{\DIFdelgraphicswidth}
\begin{picture}(1,1)
\thicklines\linethickness{2pt} 
{\color[rgb]{1,0,0}\put(0,0){\framebox(1,1){}}}
{\color[rgb]{1,0,0}\put(0,0){\line( 1,1){1}}}
{\color[rgb]{1,0,0}\put(0,1){\line(1,-1){1}}}
\end{picture}
}\hspace*{3pt}}} 
} 
\LetLtxMacro{\DIFOaddbegin}{\DIFaddbegin} 
\LetLtxMacro{\DIFOaddend}{\DIFaddend} 
\LetLtxMacro{\DIFOdelbegin}{\DIFdelbegin} 
\LetLtxMacro{\DIFOdelend}{\DIFdelend} 
\DeclareRobustCommand{\DIFaddbegin}{\DIFOaddbegin \let\includegraphics\DIFaddincludegraphics} 
\DeclareRobustCommand{\DIFaddend}{\DIFOaddend \let\includegraphics\DIFOincludegraphics} 
\DeclareRobustCommand{\DIFdelbegin}{\DIFOdelbegin \let\includegraphics\DIFdelincludegraphics} 
\DeclareRobustCommand{\DIFdelend}{\DIFOaddend \let\includegraphics\DIFOincludegraphics} 
\LetLtxMacro{\DIFOaddbeginFL}{\DIFaddbeginFL} 
\LetLtxMacro{\DIFOaddendFL}{\DIFaddendFL} 
\LetLtxMacro{\DIFOdelbeginFL}{\DIFdelbeginFL} 
\LetLtxMacro{\DIFOdelendFL}{\DIFdelendFL} 
\DeclareRobustCommand{\DIFaddbeginFL}{\DIFOaddbeginFL \let\includegraphics\DIFaddincludegraphics} 
\DeclareRobustCommand{\DIFaddendFL}{\DIFOaddendFL \let\includegraphics\DIFOincludegraphics} 
\DeclareRobustCommand{\DIFdelbeginFL}{\DIFOdelbeginFL \let\includegraphics\DIFdelincludegraphics} 
\DeclareRobustCommand{\DIFdelendFL}{\DIFOaddendFL \let\includegraphics\DIFOincludegraphics} 
\RequirePackage{listings} 
\RequirePackage{color} 
\lstdefinelanguage{DIFcode}{ 
  moredelim=[il][\color{red}\sout]{\%DIF\ <\ }, 
  moredelim=[il][\color{blue}\uwave]{\%DIF\ >\ } 
} 
\lstdefinestyle{DIFverbatimstyle}{ 
	language=DIFcode, 
	basicstyle=\ttfamily, 
	columns=fullflexible, 
	keepspaces=true 
} 
\lstnewenvironment{DIFverbatim}{\lstset{style=DIFverbatimstyle}}{} 
\lstnewenvironment{DIFverbatim*}{\lstset{style=DIFverbatimstyle,showspaces=true}}{} 

\begin{document}

\preprint{APS/123-QED}

\title{Branching Paths Statistics for confined Flows:\\
Adressing Navier-Stokes Nonlinear Transport}

\author{Daniel Yaacoub, St\'ephane Blanco, Richard Fournier and Gerjan Hagelaar}

\affiliation{
 UPS, CNRS, INPT, LAPLACE UMR CNRS 5213, Universit\'e de Toulouse,\\
118 route de Narbonne, F-31065 Toulouse, Cedex 9, France
}

\author{Jean-Fran\c cois Cornet, J\'er\'emi Dauchet, and Thomas Vourc'h}

 \affiliation{
 Universit\'e Clermont Auvergne, Clermont Auvergne INP,\\
CNRS, Institut Pascal, F-63000 Clermont-Ferrand, France
}

\date{\today}

\begin{abstract}
Recent advances have allowed to tackle exact path-space probabilistic representations of macroscopic advection-diffusion models involving advection nonlinearities by step forward approaches in terms of continuous branching stochastic processes.
Yet, the need of such paradigm shift is huge for the broad flied of fluid flows.
In deed, wherever for climate dynamics, engeenering, geophysical and planetary formations, or biomedical applications, complex transport phenomena involving diffusion and advection in confined domains set the physics.
In this work, we advance this framework by casting such branching representations within the class of Navier-Stokes strongly nonlinear transport.
This yields novel propagator representations for fluid dynamics and opens new routes for efficient simulations of fluids in confined domains by use of new Backward Monte Carlo algorithms. 
\end{abstract}


\maketitle



\section{Introduction}

\paragraph{Context. }

In many fields concerned with climate modeling, atmospheric dynamics, planetary formations, geophysical convection and tidal phenomena in planetary interiors, heat and mass transfers in combustion-related problems, fire dynamics, microfluidic cooling of electronic systems, reactive and industrial flows in process and chemical engeneering, or even crowd and traffic modeling and biomedical applications such as blood, lymph and bio-particle dynamics, complex transport phenomena in confined geometries set the physics and the challenge lies in understanding properly nonlinear advecto-diffusion.    
Besides insightful physical representations of such phenomena, the demand for robust reference solution and efficient computations is huge.


In this regard, providing both conceptual clarity and computational tools, building structures that bridge physical interpretation and computational feasibility is today a challenge uniting these communities, both theoretical and applicative.

For incompressible fluids - of viscosity $\nu$ and density $\rho$ - confined in a domain $\Omega$, the dynamic of the velocity field $\mathbf{v}$ is described by the following Navier-Stokes transport equation :
\begin{widetext}
\begin{equation}
\p_t\mathbf{v}(\mathbf{r},t)+\left(\mathbf{v}(\mathbf{r},t)\cdot\boldsymbol{\nabla}\right)\mathbf{v}(\mathbf{r},t)=\nu\boldsymbol{\nabla}^2\mathbf{v}(\mathbf{r},t)-\boldsymbol{\nabla}p(\mathbf{r},t)/\rho+\mathbf{f}(\mathbf{r},t)\label{eq:NS}
\end{equation}
\end{widetext}
along with $\boldsymbol{\nabla}\cdot\mathbf{v}(\mathbf{r},t)=0$ for all $\mathbf{r}\in\mathring{\Omega}$ and $t\in[t_\text{o};+\infty[$.
This unstationary advecto-reacto-diffusive transport equation is deterministic and lies in a strongly nonlinear Partial Differential Equation (PDE) in which the diffusive transport $\nu\boldsymbol{\nabla}^2\mathbf{v}$ is due to viscous effects and $\boldsymbol{\nabla}p/\rho+\mathbf{f}$ stands for volumic source terms due to pressure effects and eventual external volumic forcing. The advective transport term $(\mathbf{v}\cdot\boldsymbol{\nabla})\mathbf{v}$ stand for the main nonlinearity appearing in equation \eqref{eq:NS} since the velocity field $\mathbf{v}$ is localy advected by itself. The following work aims at tackling this specific nonlinearity by casting insightfull Feynman-Kac probabilistic representations within this particular nonlinearity class and constructing new statistical estimators of the velocity field based on Branching Backward Monte Carlo (BBMC) algorithms.
In the following framework, we focus on prescribed Cauchy/Dirichlet Initial Boundary Value (IBV) problem
$\mathbf{v}_\text{IBV}(\mathbf{r},t)\equiv\mathds{1}_{\{\mathbf{r}\in\p\Omega\}}\mathbf{v}^{\p\Omega}(\mathbf{r},t)
+\mathds{1}_{\{t=t_\text{o}\}}\mathbf{v}_\text{o}(\mathbf{r})$,
where $\mathbf{v}^{\p\Omega}$ stands for the boundary field and $\mathbf{v}_\text{o}$ the initial field. In cases of usual no-slip boundary conditions, the tangential component of $\mathbf{v}^{\p\Omega}$ equals the velocity of boundaries $\p\Omega$ whereas its normal component is nul.

\paragraph{Probabilistic representations of Navier-Stokes non-linear PDEs. }

From Einstein’s Brownian motion to Feynman’s path-integral picture, the dual interplay between probabilistic perspective and macroscopic deterministic continuous fields continually reshaped how physicists build intuition about transport and propagation. This dual determinisitic-probabilistic interpretation, fundamentally based on superposition and linearity, has disseminated in most fields of linear physics as for instance diffusive phenomena including heat transfers \cite{Einstein_1905,Kakutani_1944,Phillips_1923,Lewy_1928,Haji_1966,Tregan_2023}, electromagnetism \cite{Budaev_2002,Budaev_2005}, wave propagation \cite{Kac_1974,Zhang_2019}, or linear transport including neutronics and radiative transfer \cite{Maire_2006,Lejay_2010,Tregan_2023,Tessendorf_1987}, mainly because it produces flexible intuitions. 

Probabilistic representations of nonlinear Partial Differential Equations
(PDEs) have been unlocked - until recent breakthroughs - by step forward approaches extending Feynman-Kac theory, thus bringing renewed insights in terms of path-space propagative pictures. 
This has resulted in reactive nonlinearities, such as Boltzmann kinetic equation~\cite{Nyffenegger_2024,Terree_2022,Pulvirenti_2018,Kac_1956,McKean_1966,McKean_1967}, Kolmogorov-Petrovsky-Piskunov (KPP) reaction-diffusion
equations~\cite{Skorokhod_1964,McKean_1975,Ermakov_1989} or non-linear
Fredholm equations \cite{Dimov_2000} benefiting from a powerfull conceptual framework  with a unique process propagating toward sources, so-called branching stochastic process or stochastic cascade. 
Such non-linear PDEs are represented in a single path-space instead of an
infinity of inlaid ones by means of trees underlaid by branching stochastic processes.

First indroduced by \cite{Kolmogorov_1947} in 1947 (after \cite{Bienayme_1845,Galton_1875}) to let the theoretical foundations for branching Markov processes and then Continuous Branching Stochastic Processes (CBSP) \cite{jirina_1958,Lamperti_1967}, the first use of CBSP was finally made by~\cite{Skorokhod_1964,McKean_1975} to provide probabilistic representation of solutions to nonlinear PDEs.
First, branching brownian motion allowed Feynman-Kac's representation of KPP reaction-diffusion equations \cite{KPP_1937} : $\p_t\eta(\mathbf{r},t)=D\nabla^2\eta(\mathbf{r},t)+f[\eta(\mathbf{r},t)]$ in
which the non-linearity occurs within the source term $f[\eta]$ (\textit{e.g.}
Fisher-KPP: $f[\eta]=\eta(1-\eta)$~\cite{Fisher_1937}). 
Concerning Navier-Stokes equations, one has to deal with another class of nonlinearity since it occurs through the advection field, being itself the solution of the PDE.

Up to now, many probabilistic representations for free-space Navier-Stokes have
treated the nonlinear terms involving the advection field as volumic sources
~\cite{Labordere_2019,Nguwi_2023,Busnello_1999}, rather than
considering it as part of the stochastic process.
These previous works permitted thus to make use of CBSPs previously developed for KPP's reactive nonlinearities, in a similar vein as~\cite{McKean_1975}.
These approaches rely on the probabilistic representations of spatial derivatives using Malliavin stochastic calculus~\cite{Fournie_1999,Warren_2012}. 
Another approach is to study Fourier-space representations of Navier-Stokes equations. 
Thusterms involving velocity naturally become reactive nonlinearities, which also benefit from previous developments for KPP equations~\cite{Bhattacharya_2003,Ossiander_2005,LeJan_1997}.
Stochastic cascades and branching trees are take then place in the Fourier dual space.
Although these strategies have achieved a huge step forward in being able to provide probabilistic representations and propagative insights of such strongly nonlinear PDE, they remain incompatible with confined domains (especially due to the use of Malliavin calculus). This is a major issue for many applications mentioned above.

On another hand, contrasting probabilistic representations of Navier-Stokes equations compatible with confined domains have been advanced by considering nonlinear advection terms as being fully part of the process itself.
They can be conceptualized as an infinity of inlaid path-spaces~\cite{McKean_1966,Izydorczyk_2019}.
Insightfull details will be presented in section \ref{sec:II}. 
This approach has been applied to Stokes-Burger~\cite{Calderoni_1983}, or Navier-Stokes \cite{Lejay_2020} equations and subsequent statistical estimations based on these representations have been investigated either by pointwise \cite{Rioux_2022,Sugimoto_2024} or particle-systems approaches \cite{Milstein_2012}.
As we will discuss in section \ref{sec:II}, the cost is huge, since in comparison with KPP's branching trees, no path-space underlaid by a unique branching stochastic process propagating sources can be build.\\

\paragraph{The idea of CBSP for Navier-Stokes nonlinear transport and outline.}

Assume that the advection field $\mathbf{v}$ is known as the
expectation $\bb{E}[\boldsymbol{\cc{V}}]$ of a random velocity
$\boldsymbol{\cc{V}}$, that is a Feynman-Kac's representation of $\mathbf{v}$ is known for equation \eqref{eq:NS}. If in place of the advection field we were dealing with a reactive term, in the vein of Skorokhod, Mckean or Dimov 
 \cite{Skorokhod_1964,McKean_1975,Dimov_2000}, we could
replace $\mathbf{v}$ by $\boldsymbol{\cc{V}}$ in the stochastic process underlaying such a representation.
In such a reactive nonlinearity, this would be correct and the nonlinearity would exactly be represented.
However, doing so in the case of an advective nonlinearity would lead to a
spurious situation. 
How would it be possible to reconstruct such a ballistic stream line with an advective stochastic process using a random velocity that
never equals the true field value of $\mathbf{v}$ \cite{Yaacoub_2025}? 

This counterintuitive idea has prevented the use of branching stochastic processes for strongly nonlinear drift-diffusion transport in confined flows including Navier-Stokes equations, but recent breakthroughs \cite{Yaacoub_2025} have intended to show that this intuition comes from an improper limit inversion. 
With this view, we briefly transpose the recent theoretical framework extenting Feynman-Kac's therory to the nonlinear transport of velocity field described by Navier-Stokes PDE in section \ref{sec:II}.
By reconnecting such stochastic dynamics to deterministic flow descriptions, we develop then novel statistical estimators based on this new probabilistic representation within the context of backward pointwise Monte Carlo methods leading to new branching algorithms (BBMC) completely independent of the geometric desciption of considered systems.
Numerical praticability of such estimators is finally tested on specific analytical benchmarks in both unsteady situations and confined geometries : 1. Free-space unsteady Lamb-Oseen vortex, 2. Confined unsteady damped Taylor-Couette flow.

\section{Branching Path-space probabilistic representation}\label{sec:II}


\begin{widetext}
\begin{tcolorbox}[
  enhanced,
  breakable,
  width=\textwidth,
  boxrule=0.4pt,
  left=3pt,
  right=3pt,
  colback=gray!10,
  colframe=black
]
  \begin{center}
  \textbf{Path-space probabilistic representation}
  \end{center}
  
$~~~~$Feynman-Kac's framework initially aims at providing probabilistic insights into the solution of a deterministic field physics described by a linear parabolic PDE, by resorting to a probabilistic perspective and underlying stochastic processes such as the brownian motion, as it was initiated by Bachelier \cite{Bachelier_1900,Bachelier_1901}
, Einstein \cite{Einstein_1905} and Smolukowsky \cite{Smoluchowski_1906a} between 1900 and 1906 with the diffusion equation and later on by Kakutani in 1944 and 1945 \cite{Kakutani_1944,Kakutani_1945}.
Kac and Feynman advanced this framework by casting Green propagators and path integrals within measure theory, and thus defining solutions as expectations over stochastic processes for a broad class of  operators \cite{Kac_1947,Feynman_1948,Kac_1949, Kac_1951}. Feynman-Kac's probabilistic representation of the velocity field $\mathbf{v}$ submitted to Navier-Stokes equation \eqref{eq:NS} at a given probe position $(\mathbf{r},t)$ writes :
\begin{equation}
\mathbf{v}(\mathbf{r},t)=\bb{E}_{\boldsymbol{\cc{R}}_s}\Bigg[\mathbf{v}_{\text{IBV}}\big(\boldsymbol{\cc{R}}_{\cc{T}},t-\cc{T}\big)+\int_{\text{o}}^{\cc{T}}\ds s~\left(\mathbf{f}\big(\boldsymbol{\cc{R}}_s,t-s\big)-\boldsymbol{\nabla}p\big(\boldsymbol{\cc{R}}_s,t-s\big)/\rho\right)\Bigg|\boldsymbol{\cc{R}}_\text{o}=\mathbf{r}\Bigg]\equiv\bb{E}_{\boldsymbol{\cc{V}}}\left[\boldsymbol{\cc{V}}|\mathbf{r},t\right]\label{eq:FK}
\end{equation}
$\boldsymbol{\cc{R}}_\cc{T}$ is an It\^o integral~\cite{Ito_1944} defined as the
continuous limit of the sum of stochastic increments $\sum_i\delta\boldsymbol{\cc{R}}_{i\delta s}$ as $\delta s\to 0$. 
In such a limit, the stochastic process $\{\boldsymbol{\cc{R}}\}_s$ is the family of random variables which can be defined by the stochastic differential equation  $\ds\boldsymbol{\cc{R}}_s=-\mathbf{v}\left(\boldsymbol{\cc{R}}_{s},t-s\right)\ds s+\sqrt{2\nu}\ds\boldsymbol{\ff{W}}_{s}$ with $\boldsymbol{\cc{R}}_\text{o}=\mathbf{r}$ and $\ds\boldsymbol{\ff{W}}_s$ the Gaussian Wiener process.\\ 

$~~~~$If the observable represented was advected by a known and prescribed field $\mathbf{v}$, realizations of $\{\boldsymbol{\cc{R}}_s\}_s$ would describe a continuous brownian path $\{\mathbf{r}_s\}_s$ starting from $\mathbf{r}$ and backwardly propagating until a boundary/initial/volumic source is found (within the meaning of Green).
According to \eqref{eq:FK}, this observable would results in the expected value of exponentially attenuated initial/boundary/volumic sources encountered along each path. 
The first passage time of this stochastic process to the boundary $\p\Omega$ is
a random variable defined as $\cc{T}_{\p\Omega}:={\text{inf}}\{s|\boldsymbol{\cc{R}}_s\notin\mathring{\Omega}\}$.
The stopping time $\cc{T}:=\text{min}\{\cc{T}_{\p\Omega},t-t_\text{o}\}$ is
either $\cc{T}_{\p\Omega}$, in which case the Dirichlet boundary condition $\mathbf{v}^{\p\Omega}$ is taken for $\mathbf{v}_\text{IBV}$, or $\cc{T}=t-t_\text{o}$ if the initial instant is reached before the process exits the domain $\Omega$, in which case the initial condition $\mathbf{v}_\text{o}$ is taken for $\mathbf{v}_\text{IBV}$.
The ensuing set of paths would therefore draws a canonical path-space and the Feynman-Kac representation \eqref{eq:FK} could be understood as a path-integral over this
Wiener-measurable functional domain~\cite{Feynman_1948,Onsager_1953,Wiener_1921}:
\begin{equation}
\mathbf{v}(\mathbf{r},t)=\int_{\boldsymbol{\Gamma}}\cc{D}\bb{P}[\boldsymbol{\gamma}]~\left(\mathbf{v}_\text{IBV}(\boldsymbol{\gamma}(\cc{T}[\boldsymbol{\gamma}]),t-\cc{T}[\boldsymbol{\gamma}])+\int_\text{o}^{\cc{T}[\boldsymbol{\gamma}]}\ds \xi~(\mathbf{f}(t-\xi)-\boldsymbol{\nabla}p(t-\xi)/\rho)\right)
\end{equation}
given the Wiener measure $\cc{D}\bb{P}[\boldsymbol{\gamma}]$ over the path-space $\boldsymbol{\Gamma}$.\\

$~~~~$However, in the case of Navier-Stokes nonlinear transport, the stochastic processe $\{\boldsymbol{\cc{R}}_s\}_s$ depends itself on the own solution $\mathbf{v}$ to the problem since the observable and the advection field are themself the same quantity: it is a distribution-dependent process.
Two representation perspectives are hereafter exposed in this case: 1. The usual McKean representation seen as continuously inlaying the full path-space representation $\mathbf{v}=\bb{E}_{\boldsymbol{\cc{V}}}[\boldsymbol{\cc{V}}]$ within the process. 2. A recent advance allowing to recover this exact Feynman-Kac representation by using only random samples $\boldsymbol{\cc{V}}$, thus defining a continuously branching advecto-diffusive process.
\\

  \begin{minipage}[t]{0.49\textwidth}
    \textbf{1. McKean-Feynman-Kac inlaid representation.}\\
    
$~~~~$McKean representation reads as
\begin{equation}
\ds\boldsymbol{\cc{R}}_s=-\bb{E}_{\boldsymbol{\cc{V}}}\left[\boldsymbol{\cc{V}}|\boldsymbol{\cc{R}}_s,t-s\right]\ds s+\sqrt{2\nu}~\ds\boldsymbol{\ff{W}}_s\label{eq:McKean}
\end{equation}
Since equation \eqref{eq:FK} provides us with $\mathbf{v}(\mathbf{r},t)=\bb{E}_{\boldsymbol{\cc{V}}}[\boldsymbol{\cc{V}}|\boldsymbol{\cc{R}}_s,s]$, it obviously allows to recover deterministic balistic streman line.
At each time $s\in[\text{o},\cc{T}]$ the knowledge of this McKean
stochastic process $\{\boldsymbol{\cc{R}}_s\}_s$ implies
$\bb{E}_{\boldsymbol{\cc{V}}}\left[\boldsymbol{\cc{V}}|\boldsymbol{\cc{R}}_{s'},t-s'\right]$ for all $s'<s$, \textit{i.e.} the whole velocity field.
A path $\{\mathbf{r}_s\}_s$ is constructed by inlaying a full velocity
path-space centered at each $\mathbf{r}_{s'}$, drawing then, an infinite tree of inlaid path-spaces.
The cost is huge since statistical estimations based on this formulation either by particle-systems approaches \cite{Milstein_2012} or by recent pointwise Monte Carlo methods \cite{Rioux_2022,Sugimoto_2024} developped for images synthesis present a computational time explosion besides the loss of being able to define a unique branching stochastic process.

  \end{minipage}\hfill
  \begin{minipage}[t]{0.49\textwidth}
    \textbf{2. Coupled Feynman-Kac representation.}\\

$~~~~$The recent proposition made by \cite{Yaacoub_2025} reads as
\begin{equation}
\ds\boldsymbol{\widetilde{\cc{R}}}_s=-(\boldsymbol{\cc{V}}|\boldsymbol{\widetilde{\cc{R}}}_s,t-s)\ds s+\sqrt{2\nu}~\ds\boldsymbol{\ff{W}}_s\label{eq:CBSP}
\end{equation}

$~~~~$At each time $s\in[\text{o},\cc{T}]$ the knowledge of the process
$\{\widetilde{\boldsymbol{\cc{R}}}_{s}\}_{s}$ is now entirely determined by
$\boldsymbol{\cc{V}}|\widetilde{\boldsymbol{\cc{R}}}_{s'},t-s'$ for all $s'<s$,
that is the statistics of $\boldsymbol{\cc{V}}$ only, in contrast with the full
velocity field that was required above, and unknown since it is the own solution of the problem. 
A path $\{\tilde{\mathbf{r}}_s\}_s$ is constructed by embedding a unique path 
of $\boldsymbol{\cc{V}}$ centered at each $\mathbf{r}_{s'}$.
In other words, velocity paths pass on all the information about the self-coupling to the velocity model, without continuously inlaying a full path-space but drawing instead a unique branch, branching then in a stochastic cascade as for Boltzmann and KPP reactive nonlinearities representations.
$\{\widetilde{\boldsymbol{\cc{R}}}_s\}_s$ can therefore be understood as an
embedded process that includes the statistics of $\boldsymbol{\cc{V}}$.
  \end{minipage}
\textcolor{gray!10}{.}\\  \vspace{0.1cm}

$~~~~$To conclude with this formal section, it can be shown that equation \eqref{eq:FK} can be expressed as
  \begin{equation}
\mathbf{v}(\mathbf{r},t)=\bb{E}_{\boldsymbol{\cc{R}}_s,\cc{S}}\Bigg[\mathbf{v}_{\text{IBV}}\big(\boldsymbol{\cc{R}}_{\cc{T}},t-\cc{T}\big)+\frac{\mathbf{f}\big(\boldsymbol{\cc{R}}_\cc{S},t-\cc{S}\big)-\boldsymbol{\nabla}p\big(\boldsymbol{\cc{R}}_\cc{S},t-\cc{S}\big)/\rho}{p_\cc{S}(\cc{S})}\Bigg|\boldsymbol{\cc{R}}_\text{o}=\mathbf{r}\Bigg]\label{eq:FK_source}
\end{equation}
introducing an importance probability distribution function $p_\cc{S}$ of the random variable $\cc{S}$ disctributed on $[\text{o},\cc{T}]$.
The later formulation will be usefull in the following since it wil allow to sample volumic source terms once instead of cumulating them all along each path.
\end{tcolorbox}
\end{widetext}

\section{Monte Carlo method and statistical estimators}

\paragraph{Monte Carlo algorithm.}

Starting now from the probabilistic representation \eqref{eq:FK_source}-\eqref{eq:CBSP}, the Monte Carlo method allows us to build the following statistical estimator 
\begin{equation}
\widehat{\mathbf{V}}_N(\mathbf{r},t)=\frac{1}{N}\sum_{i\in\llbracket1;N\rrbracket}(\boldsymbol{\cc{V}}_i|\mathbf{r},t)\label{eq:estimator}
\end{equation}
based on the family of independent random variables $\{\boldsymbol{\cc{V}}_i\}_{i\in\llbracket 1;N\rrbracket}$ of realizations $\mathbf{v}_i$ and identically distributed with rescpect to $\boldsymbol{\cc{V}}$.
As the number $N$ of samples tend to infinity, $\widehat{\mathbf{v}}_N(\mathbf{r},t)$ converges in probability toward $\mathbf{v}(\mathbf{r},t)$ by the law of large numbers.
Algorithm \ref{alg:MC} depict how to build a statistical estimation of $\mathbf{v}$ at a given probe position $(\mathbf{r},t)$ by use of the statistical estimator \eqref{eq:estimator}.

\begin{algorithm}[H]
\label{alg:MC}
  \caption{Single Branching path-space Monte Carlo}
  \label{}
   $\bullet$ Number of realisations : $N$\;
   $\bullet$ Probe position : $(\mathbf{r},t)$\;
   $\bullet$ Initialisation : $i=0$, $\Sigma=0$, $\Sigma_2=0$\;
   \While{$i<N$}{
   $\bullet$ $\mathbf{v}_i\leftarrow$ Sample a velocity random variable $\boldsymbol{\cc{V}}_i$ starting at $(\mathbf{r},t)$ according to Alg. \ref{alg:BSP}\;
   $\bullet$ $\Sigma\leftarrow\Sigma+\mathbf{v}_i$\;
   $\bullet$ $\Sigma_2\leftarrow\Sigma_2+\mathbf{v}_i^2$\;
   $\bullet$ $i\leftarrow i+1$\;}
   \Return Statistiacal estimation: $\Sigma/N$\\
   \Return Statistical standard deviation: $\sqrt{(\Sigma_2/N-(\Sigma/N)^2)/(N-1)}$
\end{algorithm}
\textcolor{white}{.}
\paragraph{Velocity path sampling.} 

Sampling the random variable $\boldsymbol{\cc{V}}$ implies the ability to construct the path described by the stochastic process $\ds\boldsymbol{\widetilde{\cc{R}}}_s=-(\boldsymbol{\cc{V}}|\boldsymbol{\widetilde{\cc{R}}}_s,t-s)\ds s+\sqrt{2\nu}~\ds\boldsymbol{\ff{W}}_s$ starting at $(\mathbf{r},t)$.
As $\boldsymbol{\cc{V}}$ appears itself on the definition of the branching path, it is algorithmically translated in Alg. \ref{alg:BSP} by a recursive structure.
Branching velocity paths are sampled using Maruyama's discretization scheme corresponding to a left-side Euler scheme of this stochastic differential equation. 
Defining $n$ such that $\cc{T}=n\delta s$ and providing us with a regular
subdivision $\{i\delta s|i\in\llbracket0;n-1\rrbracket\}$ of
$[\text{o};\cc{T}]$, this scheme writes
\begin{equation}
\boldsymbol{\widehat{\widetilde{\cc{R}}}}_{(i+1)\delta s}=\boldsymbol{\widehat{\widetilde{\cc{R}}}}_{i\delta
s}-\boldsymbol{\cc{V}}(\boldsymbol{\widehat{\widetilde{\cc{R}}}}_{i\delta s},t-(i+1)\delta s)\delta
s+\sqrt{2D}\delta\boldsymbol{\ff{W}}_{i\delta s}
\end{equation}
if we choose a right-side discretisation for the purely temporal argument of $\boldsymbol{\cc{V}}$.
The fundamental Wiener increment 
$\delta\boldsymbol{\ff{W}}_{i\delta s}$ is a gaussian vector with mean
$\bb{E}[\delta\boldsymbol{\ff{W}}_{i\delta s}]=\boldsymbol{0}$ and variance
$\bb{V}[\delta\boldsymbol{\ff{W}}_{i\delta s}]=2D\delta s\delta_{j,k}$ ($(j,k)$
standing for component labels). 
The continuous limit is obtained when $N\to\infty$, that is $\delta s\to0$, and
has to be understood as convergence in probability in the sense of Ito.
In this limit $\{\boldsymbol{\widehat{\widetilde{\cc{R}}}}_s\}_s$ tend to $\{\boldsymbol{\widetilde{\cc{R}}}_s\}_s$. 
Alg. \ref{alg:BSP} presents the sampling method for $(\boldsymbol{\cc{V}}|\mathbf{r},t)$.

\begin{algorithm}[H]
\label{alg:BSP}
  \SetAlgoVlined
  \caption{Velocity path sampling : $(\boldsymbol{\cc{V}}\mid\mathbf{r},t)$}
  $\bullet$ Initial probe position: $\hat{\mathbf{r}}=\mathbf{r}$\;
  $\bullet$ Initial path time: $s=0$\;
  $\bullet$ Path discretization time: $\delta s$\;
  $\bullet$ Weights $\mathbf{w}_{\mathbf{v}}=\mathbf{0}$ and $\mathbf{w}_{\mathbf{F}}=\mathbf{0}$\;
  \textbf{exit} = False\;
  \While{\textbf{exit} $\neq$ True}{
    $\bullet$ $s \leftarrow s + \delta s$\;
    $\bullet$ $\mathbf{v} \leftarrow$ Sample $\bigl(\boldsymbol{\mathcal{V}}\mid \hat{\mathbf{r}},\,t-s\bigr)$ according to Alg. \ref{alg:BSP}\;
    $\bullet$ $\delta\mathbf{w} \leftarrow$ Sample $\delta\boldsymbol{W}_s$ according to $\mathcal{N}(0,2\nu\delta s)$\;
    $\bullet$ $\hat{\mathbf{r}} \leftarrow \hat{\mathbf{r}} - \mathbf{v}\,\delta s + \delta\mathbf{w}$\;
  \If{$\hat{\mathbf{r}}\notin\mathring{\Omega}$}{
    $\bullet$ percolation position $\hat{\mathbf{r}}_{\partial\Omega}$ and time $t_{\partial\Omega}$ to the boundary obtained by linear intersection between $[\hat{\mathbf{r}},~\hat{\mathbf{r}}-\mathbf{v}\delta s+\delta\mathbf{w}]$ segment and $\partial\Omega$\;
    $\bullet$ $\mathbf{w}_{\mathbf{v}} = \mathbf{v}^{\partial\Omega}\bigl(\hat{\mathbf{r}}_{\partial\Omega},\,t-t_{\partial\Omega}\bigr)$\;
    $\bullet$ \textbf{exit} = True\;
  }
  \If{$s \ge t-t_\text{o}$}{
    $\bullet$ $\mathbf{w}_{\mathbf{v}} = v_{\mathrm{o}}(\hat{\mathbf{r}})$\;
    $\bullet$ \textbf{exit} = True\;
  }}
  $\bullet$ $s_{\text{rand}} \leftarrow$ Sample $\mathcal{S}$ according to $p_{\mathcal{S}}$ on $[0,s]$\;
  $\bullet$ Evaluate $p_{\mathcal{S}}(s_{\text{rand}})$\;
  $\bullet$ $\mathbf{w}_{\mathbf{F}} = (\mathbf{f}(\hat{\mathbf{r}}_{s_\text{rand}},t-s_\text{rand})-\boldsymbol{\nabla}p(\hat{\mathbf{r}}_{s_\text{rand}},t-s_\text{rand})/\rho)/p_{\mathcal{S}}(s_{\text{rand}})$\;
  \Return $\mathbf{w}_{\mathbf{v}} + \mathbf{w}_{\mathbf{F}}$\;
\end{algorithm}

\begin{figure}[H]
\begin{center}
\begin{minipage}[b]{0.45\linewidth}
    a)\includegraphics[width = 1\linewidth]{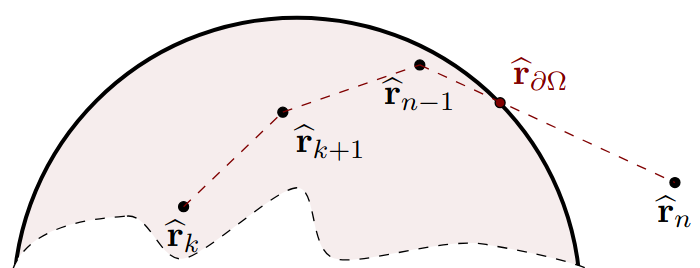}
\end{minipage}
\hspace{0.2cm}
\begin{minipage}[b]{0.45\linewidth}
    b)\includegraphics[width = 1\linewidth]{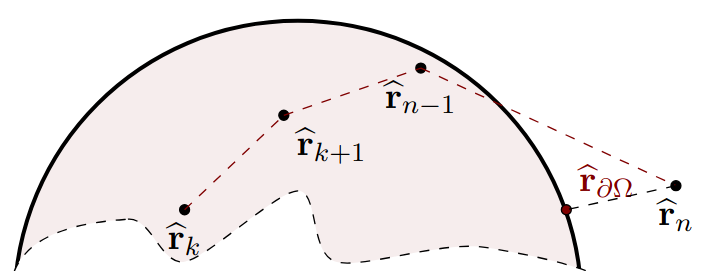}
\end{minipage}
\end{center}
	\caption{a) Ray tracing intersection with $\p\Omega$ by linear interpolation between the latest sampled position in $\mathring{\Omega}$ and the first sampled position in $\bb{R}^3\backslash\overline{\Omega}$. b) First passage percolation position infered by the nearest orthogonal projection.} 
	\label{fig:percolation}
\end{figure}

$~~~~$Along with Alg. \ref{alg:MC}, Alg. \ref{alg:BSP} provides a statistical sampling of the unique path-space underlaying the exact probabilistic representation \eqref{eq:FK}-\eqref{eq:CBSP}.
The corresponding paths are branching ones but not paths inlaid with a full path-space, as it would be for McKean representation \eqref{eq:FK_source}-\eqref{eq:McKean}.
The latter would lead to nesting a Monte Carlo estimations within Monte Carlo estimations, as it is done by recent works in the community of computer graphics \cite{Rioux_2022,Sugimoto_2024}.
Concerning first passages to the boundary, two main perspectives illustrated in Fig. \ref{fig:percolation} hold for infering the first passage poristion $\hat{\mathbf{r}}_{\p\Omega}$ to the boundary $\p\Omega$: $\hat{\mathbf{r}}_{\p\Omega}\in\cc{D}\cap\p\Omega$.
The first strategy consists in linearly interpolating between the last position $\widehat{\mathbf{r}}_{n-1}$ sampled in the domain and the first position $\widehat{\mathbf{r}}_{n}$ sampled outside. $\widehat{\mathbf{r}}_{\p\Omega}\in\cc{D}\cap\p\Omega$ results then in the intersection between the straight line $\cc{D}$
\begin{equation}
\cc{D}~:~~~~\widehat{\mathbf{r}}_\cc{D}=\widehat{\mathbf{r}}_{n-1}+\sigma\frac{\widehat{\mathbf{r}}_{n}-\widehat{\mathbf{r}}_{n-1}}{||\widehat{\mathbf{r}}_{n}-\widehat{\mathbf{r}}_{n-1}||}~~~~~~~~;~\sigma\in\bb{R}_+^\star
\end{equation}
along $(\widehat{\mathbf{r}}_{n-1},\widehat{\mathbf{r}}_{n})$ and the boundary $\p\Omega$. 
This strategy allows us to infer the first passage position to the boundary only needing line/surface intersections.
By denoting $|d_{\p\Omega}(\mathbf{r})|=\underset{\mathbf{r'}\in\p\Omega}{\text{inf}}||\mathbf{r}-\mathbf{r'}||$ the distance to the nearest boundary, one can show that
\begin{equation}
\widehat{\mathbf{r}}_{\p\Omega}=\widehat{\mathbf{r}}_{n}-d_{\p\Omega}\left(\widehat{\mathbf{r}}_{n}\right)\frac{\boldsymbol{\nabla}d_{\p\Omega}\left(\widehat{\mathbf{r}}_{n}\right)}{||\boldsymbol{\nabla}d_{\p\Omega}\left(\widehat{\mathbf{r}}_{n}\right)||}
\end{equation}
since $-\boldsymbol{\nabla}d_{\p\Omega}\left(\widehat{\mathbf{r}}_{n}\right)$ indicates the direction of the nearest intersection.
This method allows to infer the first passage position to the boundary by use of surface/surface intersections. 

In the view of taking advantages of acceleration techniques developped in images synthesis and casting our work into promising frameworks opened by the computer graphics community in tackling complex geometries \cite{Sawhney_2022,Sawhney_2023,Miller_2023}, first passage positions are hereafter infered by line/surface intersections.

$~~~~$ Finally, our BBMC stastistical estimation procedure benefits from all the power of usual Monte Carlo algorithms.
First, as $\delta s\to 0$, the statistical estimator \eqref{eq:estimator} displays a null systematic error compared to the mathematical probabilistic representation and the underlaying physical model, and comes with confidence intervals.
Secondly, this approach is meshless since there is a complete orthogonality between the calculus and the description of the geometry,as illustrated line 11 of Alg. \ref{alg:BSP}: there is thus no need to discretize the space nor the time.
This last remark allow us to affirm that solving problems involving complex geometries yields no conceptual difference nor technical bottleneck, as shown in \cite{Ibarrart_2025, Villefranque_2022,Nyffenegger_2024,Bati_2023}.
Then, this approach allows one to calculate sensitivities from within the main simulation, and parallelization is straightforward.
Finally, this method is a pointwise method avoiding us from computing the whole velocity field or having to follow numerous particles interacting which each other.

\section{Results and discussions}

\begin{figure}[H]
	\centering
	{\includegraphics[width=9cm, keepaspectratio]{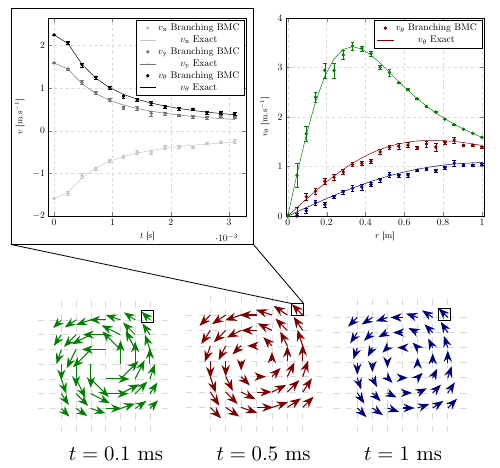}}
	\caption{Temporal and spatial profiles of the velocity field. Each Branching Backward Monte Carlo estimation is computed for $N=7\times 10^3$ samples, $\nu=2,2\times 10^2$ [m$^2$s$^{-1}$], $\Gamma=1\times 10^1$ [m$^2$s$^{-1}$] and $\rho=1\times 10^3$ [kg.m$^{-3}$]. Branching paths are sampled by $\delta s=1,7\times 10^{-5}$ [s] for $t=1\times 10^{-4}$ [s], by $\delta s=6\times 10^{-5}$ [s] for $t=5\times 10^{-4}$ [s] and by $\delta s=1,2\times 10^{-4}$ [s] for $t=1\times 10^{-3}$ [s].} 
	\label{fig:LO}
\end{figure}

\paragraph{Free-space unsteady Lamb-Oseen vortex.} We consider the velocity field $\mathbf{v}$ satisfying the incompressible condition $\boldsymbol{\nabla}\cdot\mathbf{v}=0$ and submitted to Navier-Stokes equation \eqref{eq:NS} with $\boldsymbol{\nabla}p(\mathbf{r},t)=\rho\Gamma\big(1-\text{e}^{-r^2/4\nu (t-t_\text{o})}\big)^2/2\pi r^2 \mathbf{\hat{e}_r}$ and $\mathbf{f}=\mathbf{0}$ for $\mathbf{r}\in\bb{R}^2$ and $t>t_\text{o}$. At the inital time $t_\text{o}$, $\mathbf{v}(\mathbf{r},t_\text{o})$ is imposed by the free-space Lamb-Oseen vortex $\mathbf{v}_\text{LO}(\mathbf{r},t_\text{0})=\Gamma/2\pi r$, so that one will be able to compare our estimation of $\mathbf{v}$ to the exact solution of this Cauchy problem for all $\mathbf{r}\in\bb{R}^2$ and $t>t_\text{o}$ : $\mathbf{v}(\mathbf{r},t)=\mathbf{v}_\text{LO}(\mathbf{r},t)$, given 
\begin{equation}
\mathbf{v}_\text{LO}(\mathbf{r},t)=\Gamma\big(1-\text{e}^{-r^2/4\nu (t-t_\text{o})}\big)/2\pi r\mathbf{\hat{e}_\theta}
\end{equation}
Fig. \ref{fig:LO} illustrates statistical estimations of $\mathbf{v}$ by use of our BBMC algorithm \ref{alg:MC}-\ref{alg:BSP} in the case of this 2d free-space Lamb-Oseen vortex. Both radial and temporal profiles are hereafter exposed. 

\paragraph{Confined unsteady damped Taylor-Couette flow.} 

\begin{figure}[H]
	\centering
	{\includegraphics[width=9cm, keepaspectratio]{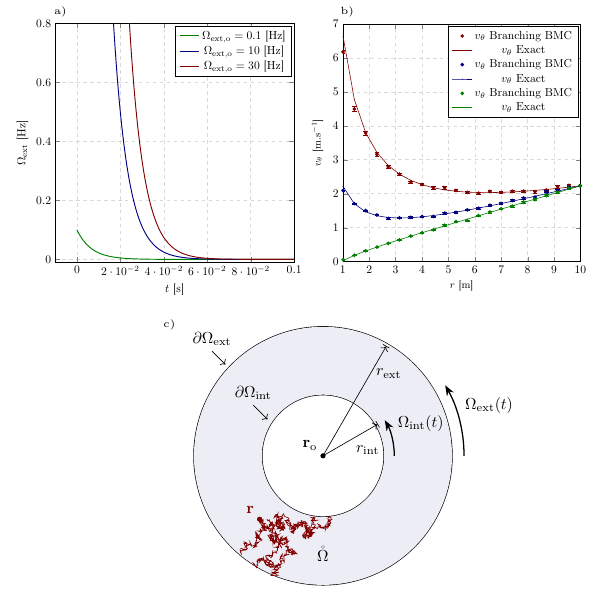}}
	\caption{Temporal profile of the angular frequencies and spatial profiles of the velocity field. Each Branching Backward Monte Carlo estimation is computed for $N=1\times 10^4$, $\nu=1\times 10^1$ [m$^2$s$^{-1}$], $R1=1$ [m], $\Omega_{2,\text{o}}=1$ [Hz], $\rho=1.10^3$ [kg.m$^{-3}$], $\lambda=1,5\times 10^2$ [Hz] and $\delta s=6\times 10^{-4}$ [s].} 
	\label{fig:TC}
\end{figure}

Circular Taylor-Couette flows have wide applications ranging from desalination to magnetohydrodynamics and also in viscosimetric analysis. A fluid of density $\rho$ and dynamic viscosity $\nu$ is confined between to rotating circles $\p\Omega_\text{int}\equiv\bb{S}^1(r_\text{int})$ and $\p\Omega_\text{ext}\equiv\bb{S}^1(r_\text{ext})$ of respective radia $r_\text{int}$ and $r_\text{ext}$ satisfying $r_\text{int}<r_\text{ext}$.
In the usual 2d Taylor-Couette flow, angular rotation frequencies of the inner and outer boundaries are prescribed and do not depend on time.
In the following exemple, such frequencies $\Omega_\text{int}$ and $\Omega_\text{ext}$ are still prescribed but depend now on time and evolve such as $\Omega_\text{int}(t)=\Omega_{\text{int},\text{o}}\text{e}^{-\lambda t}$ and $\Omega_\text{ext}(t)=\Omega_{\text{ext},\text{o}}\text{e}^{-\lambda t}$,considering the damping parameter $\lambda$.
We consider the velocity field $\mathbf{v}$ satisfying the incompressible condition $\boldsymbol{\nabla}\cdot\mathbf{v}=0$ and submitted to Navier-Stokes equation () with $\boldsymbol{\nabla}p(\mathbf{r},t)=-(\rho\text{e}^{-2\lambda t}((\Omega_{\text{int},\text{o}}r_\text{int}^2-\Omega_{\text{ext},\text{o}}r_\text{ext}^2))r/(r^2_\text{int}-r_\text{ext}^2)+(\Omega_{\text{ext},\text{o}}-\Omega_{\text{int},\text{o}})r_\text{int}^2r_\text{ext}^2/((r_\text{int}^2-r_\text{ext}^2)r))^2/r)\mathbf{\hat{e}_r}$ and $\mathbf{f}(\mathbf{r},t)=-\lambda\text{e}^{-\lambda t}((\Omega_{\text{int},\text{o}}r_\text{int}^2-\Omega_{\text{ext},\text{o}}r_\text{ext}^2))r/(r^2_\text{int}-r_\text{ext}^2)+(\Omega_{\text{ext},\text{o}}-\Omega_{\text{int},\text{o}})r_\text{int}^2r_\text{ext}^2/((r_\text{int}^2-r_\text{ext}^2)r))\mathbf{\hat{e}_\theta}$ for $\mathbf{r}\in\mathring{\Omega}\equiv\bb{B}^2(r_\text{ext})\backslash\bb{B}^2(r_\text{int})$. At the boundary $\p\Omega\equiv\p\Omega_\text{int}\cup\p\Omega_\text{ext}$, no-slip conditions impose $\mathbf{v}(\mathbf{r}\in\p\Omega,t)\cdot\mathbf{\hat{e}_\theta}=r_\text{int}\Omega_{\text{int}}(t)\mathds{1}_{\{\mathbf{r}\in\p\Omega_\text{int}\}}+r_\text{ext}\Omega_{\text{ext}}(t)\mathds{1}_{\{\mathbf{r}\in\p\Omega_\text{ext}\}}$ are considered. Finally, the initiale condition is fixed by the usual Taylor-Couette profile $\mathbf{v}(\mathbf{r},t_\text{o})=((\Omega_{\text{int},\text{o}}r_\text{int}^2-\Omega_{\text{ext},\text{o}}r_\text{ext}^2))r/(r^2_\text{int}-r_\text{ext}^2)+(\Omega_{\text{ext},\text{o}}-\Omega_{\text{int},\text{o}})r_\text{int}^2r_\text{ext}^2/((r_\text{int}^2-r_\text{ext}^2)r))\mathbf{\hat{e}_\theta}$.
In this case, we are able to compare our nimerical estimations of the velocity  to the eaxact analytical solution of this Cauchy-Dirichlet Initial-Boundary Value Problem :
\begin{equation}
\mathbf{v}_\text{TC}(\mathbf{r})=\text{e}^{-\lambda t}\left(\alpha_\text{o}r+\frac{\beta_\text{o}}{r}\right)\mathbf{\hat{e}_\theta}
\end{equation}
given 
\begin{equation}
\alpha_\text{o}=\frac{\Omega_{\text{int},\text{o}}r_\text{int}^2-\Omega_{\text{ext},\text{o}}r_\text{ext}^2}{r^2_\text{int}-r_\text{ext}^2}
\end{equation}
and
\begin{equation}\beta_\text{o}=\frac{(\Omega_{\text{ext},\text{o}}-\Omega_{\text{int},\text{o}})r_\text{int}^2r_\text{ext}^2}{r_\text{int}^2-r_\text{ext}^2}
\end{equation}
noting $r=||\mathbf{r}||$.
Fig. \ref{fig:TC} illustrates statistical estimations of $\mathbf{v}$ by use of our BBMC algorithm \ref{alg:MC}-\ref{alg:BSP} for three various damping regimes.

\section{conclusions and perspectives}

In the present work, we have advanced recent probabilistic approaches of nonlinear advecto-reacto-diffusive transport to the particular class of fluid flows described by incompressible Navier-Stokes equations in confined domains.
Our formulation shows how expectations over a single, well-defined branching path-space recover deterministic flow maps.
Taken together, these results bridge physical interpretation and computational feasibility across scientific communities concened with fluid flows and nonlinear transport phenomena in confined domains and offer a new descriptive framework

Wherever fluid phenomena, geometric sophistication, and the demand for robust reference solutions, impose stringent limits (whether in advanced engineering systems, intricate biomedical models, or climate modeling), the presented framework delivers a promising perspective.
By decoupling computational effort from the system’s inherent complexity while maintaining rigorous probabilistic foundations, it lays the groundwork for tackling numerous challenges long deemed out of reach, fundamentally redefining standards of predictive power and scientific interpretation of fluid flows.

This work immediately unfolds along two crucial dimensions. On the interpretative front, it fundamentally reshapes our understanding of these phenomena in terms of nonlinear propagators.
On the computational side, it opens the door to harnessing recent breakthroughs in image synthesis, yielding algorithms whose costs are remarkably insensitive to the geometric and temporal intricacies of the underlying system.
In this regard, it would be interesting to improve our Monte Carlo algorithms on large-scale systems and complex geometries since it beneficies directly from computer graphics techniques used in images synthesis and proved to be powerfull in complex physics systems \cite{Ibarrart_2025, Villefranque_2022,Nyffenegger_2024,Bati_2023}.

This work opens new routes for path-space multiphysics coupling involving fluid dynamics, until now treated with deterministic methods resultinf in  statistical/deterministic coupled algorithms. 
  
In the same vein as for nonlinear Boltzmann kinetic transport, the underlaying path-space probabilisic representation and subsequent statistical estimators involve \textit{a priori} unbound branching tree depths.
In this regard, it would be useful to explore recent advances allowing trees truncations. Such method is known as Picard series expansion and have allowed to extend the faisability of such Monte Carlo algorithms in gaz kinetics.


\begin{acknowledgments}
This work was supported by the MCMET project (ANR-23-CE46-0002) 
of the French National Research Agency
(ANR).
\end{acknowledgments}

\bibliography{biblio_merge}

@PREAMBLE{
 "\providecommand{\noopsort}[1]{}" 
 # "\providecommand{\singleletter}[1]{#1}%" 
}

@article{Miller_2023,
author = {Miller, B. and Sawhney, R. and Crane, K. and Gkioulekas, I.},
title = {Boundary Value Caching for Walk on Spheres},
year = {2023},
issue_date = {August 2023},
publisher = {Association for Computing Machinery},
address = {New York, NY, USA},
volume = {42},
number = {4},
issn = {0730-0301},
url = {https://doi.org/10.1145/3592400},
doi = {10.1145/3592400},
journal = {ACM Trans. Graph.},
month = jul,
articleno = {82},
numpages = {11}
}

@article{Sawhney_2023,
author = {Sawhney, R. and Miller, B. and Gkioulekas, I. and Crane, K.},
title = {Walk on Stars: A Grid-Free Monte Carlo Method for PDEs with Neumann Boundary Conditions},
year = {2023},
issue_date = {August 2023},
publisher = {Association for Computing Machinery},
address = {New York, NY, USA},
volume = {42},
number = {4},
issn = {0730-0301},
url = {https://doi.org/10.1145/3592398},
doi = {10.1145/3592398},
journal = {ACM Trans. Graph.},
month = jul,
articleno = {80},
numpages = {20}
}

@article{Sawhney_2022,
author = {Sawhney, R. and Seyb, D. and Jarosz, W. and Crane, K.},
title = {Grid-free Monte Carlo for PDEs with spatially varying coefficients},
year = {2022},
issue_date = {July 2022},
publisher = {Association for Computing Machinery},
address = {New York, NY, USA},
volume = {41},
number = {4},
issn = {0730-0301},
url = {https://doi.org/10.1145/3528223.3530134},
doi = {10.1145/3528223.3530134},
journal = {ACM Trans. Graph.},
month = jul,
articleno = {53},
numpages = {17}
}

@article{Pulvirenti_2018,

title = {Microscopic solutions of the Boltzmann-Enskog equation in the series representation},

journal = {Kinetic and Related Models},

volume = {11},

number = {4},

pages = {911-931}
,
year = {2018},

issn = {1937-5093},

doi = {10.3934/krm.2018036},

url = {https://www.aimsciences.org/article/id/2732f857-4cd2-44d6-8ce8-515bfa2f636d},

author = {Pulvirenti, M. and Simonella, S. and Trushechkin, A.}

}

@article{Bati_2023,
author = {Bati, M. and Blanco, S. and Coustet, C. and Eymet, V. and Forest, V. and Fournier, R. and Gautrais, J. and Mellado, N. and Paulin, M. and Piaud, B.},
title = {Coupling Conduction, Convection and Radiative Transfer in a Single Path-Space: Application to Infrared Rendering},
year = {2023},
issue_date = {August 2023},
publisher = {Association for Computing Machinery},
address = {New York, NY, USA},
volume = {42},
number = {4},
issn = {0730-0301},
url = {https://doi.org/10.1145/3592121},
doi = {10.1145/3592121},
journal = {ACM Trans. Graph.},
month = jul,
articleno = {79},
numpages = {20}
}

@article{
Villefranque_2022,
author = {Villefranque, N.  and Hourdin, F.  and d’Alençon, L.  and Blanco, S.  and Boucher, O.  and Caliot, C.  and Coustet, C.  and Dauchet, J.  and El Hafi, M.  and Eymet, V.  and Farges, O.  and Forest, V.  and Fournier, R.  and Gautrais, J.  and Masson, V.  and Piaud, B.  and Schoetter, R.},
title = {The “teapot in a city”: A paradigm shift in urban climate modeling},
journal = {Science Advances},
volume = {8},
number = {27},
pages = {eabp8934},
year = {2022},
doi = {10.1126/sciadv.abp8934},
URL = {https://www.science.org/doi/abs/10.1126/sciadv.abp8934},
eprint = {https://www.science.org/doi/pdf/10.1126/sciadv.abp8934}}

@article{Ibarrart_2025,
    doi = {10.1371/journal.pone.0330604},
    author = {Ibarrart, Loris AND Blanco, Stéphane AND Caliot, Cyril AND Dauchet, Jérémi AND Eibner, Simon AND El Hafi, Mouna AND Farges, Olivier AND Forest, Vincent AND Fournier, Richard AND Gautrais, Jacques AND Konduru, Raj AND Penazzi, Léa AND Trégan, Jean-Marc AND Vourc’h, Thomas AND Yaacoub, Daniel},
    journal = {PLOS ONE},
    publisher = {Public Library of Science},
    title = {Advection, diffusion and linear transport in a single path-sampling Monte-Carlo algorithm: Getting insensitive to geometrical refinement},
    year = {2025},
    month = {09},
    volume = {20},
    url = {https://doi.org/10.1371/journal.pone.0330604},
    pages = {1-33},
    number = {9},

}

@article{
Nyffenegger_2024,
author = {Nyffenegger-Péré, Y.  and Armante, R.  and Bati, M.  and Blanco, S.  and Dufresne, J.-L.  and El Hafi, M.  and Eymet, V.  and Forest, V.  and  Fournier, R.  and Gautrais, J.  and Lebrun, R.  and Mellado, N.  and Mourtaday, N.  and Paulin, M.},
title = {Spectrally refined unbiased Monte Carlo estimate of the Earth’s global radiative cooling},
journal = {Proceedings of the National Academy of Sciences},
volume = {121},
number = {5},
pages = {e2315492121},
year = {2024},
doi = {10.1073/pnas.2315492121},
URL = {https://www.pnas.org/doi/abs/10.1073/pnas.2315492121},
eprint = {https://www.pnas.org/doi/pdf/10.1073/pnas.2315492121}}

@article{Tregan_2023,
    doi = {10.1371/journal.pone.0283681},
    author = {Tregan, J.-M. and Amestoy, J.-L. and Bati, M. and Bezian, J.-J. and Blanco, S. and Brunel, L. and Caliot, C. and Charon, J. and Cornet, J.-F. and Coustet, C. and d’Alençon, L. and Dauchet, J. and Dutour, S. and Eibner, S. and El Hafi, M. and Eymet, V. and Farges, O. and Forest, V. and Fournier, R. and Galtier, M. and Gattepaille, V. and Gautrais, J. and He, Z. AND Hourdin, F. and Ibarrart, L. and Joly, J.-L. and Lapeyre, P. and Lavieille, P. and Lecureux, M.-H. and Lluc, J. and Miscevic, M. and Mourtaday, N. and Nyffenegger-Péré, Y. and Pelissier, L. and Penazzi, L. and Piaud, B. and Rodrigues-Viguier, C. and Roques, G. and Roger, M. and Saez, T. and Terrée, G. and Villefranque, N. and Vourc’h, T. and Yaacoub, D.},
    journal = {PLOS ONE},
    publisher = {Public Library of Science},
    title = {Coupling radiative, conductive and convective heat-transfers in a single Monte Carlo algorithm: A general theoretical framework for linear situations},
    year = {2023},
    month = {04},
    volume = {18},
    url = {https://doi.org/10.1371/journal.pone.0283681},
    pages = {1-54},
    number = {4},

}

@article{Budaev_2002,
  title={Application of random walk methods to wave propagation},
  author={Budaev, B. V. and Bogy, D. B.},
  journal={Quarterly Journal of Mechanics and Applied Mathematics},
  volume={55},
  number={2},
  pages={209--226},
  year={2002},
  publisher={OUP}
}

@article{Wiener_1921,
 ISSN = {00278424, 10916490},
 URL = {http://www.jstor.org/stable/84448},
 author = {Wiener, N.},
 journal = {Proceedings of the National Academy of Sciences of the United States of America},
 number = {9},
 pages = {253--260},
 publisher = {National Academy of Sciences},
 title = {The Average of an Analytic Functional},
 urldate = {2024-10-24},
 volume = {7},
 year = {1921}
}

@article{Tessendorf_1987,
  title = {Radiative transfer as a sum over paths},
  author = {Tessendorf, J.},
  journal = {Phys. Rev. A},
  volume = {35},
  issue = {2},
  pages = {872--878},
  numpages = {0},
  year = {1987},
  month = {Jan},
  publisher = {American Physical Society},
  doi = {10.1103/PhysRevA.35.872},
  url = {https://link.aps.org/doi/10.1103/PhysRevA.35.872}
}

@article{Onsager_1953,
  title = {Fluctuations and Irreversible Processes},
  author = {Onsager, L. and Machlup, S.},
  journal = {Phys. Rev.},
  volume = {91},
  issue = {6},
  pages = {1505--1512},
  numpages = {0},
  year = {1953},
  month = {Sep},
  publisher = {American Physical Society},
  doi = {10.1103/PhysRev.91.1505},
  url = {https://link.aps.org/doi/10.1103/PhysRev.91.1505}
}

@ARTICLE{Einstein_1905,
       author = {Einstein, A.},
        title = {{\"U}ber die von der molekularkinetischen Theorie der W{\"a}rme geforderte Bewegung von in ruhenden Fl{\"u}ssigkeiten suspendierten Teilchen},
      journal = {Annalen der Physik},
         year = 1905,
        month = jan,
       volume = {322},
       number = {8},
        pages = {549-560},
          doi = {10.1002/andp.19053220806},
       adsurl = {https://ui.adsabs.harvard.edu/abs/1905AnP...322..549E}
}

@article{Smoluchowski_1906a,
  title = {Essai d’une théorie du mouvement brownien et de milieux troubles},
  author = {Smoluchowski, M.},
  journal = {Bull. Acad. Sci. Cracovie},
  pages = {577–602},
  year = {1906}
}

@article{Kac_1949,
 ISSN = {00029947, 10886850},
 URL = {http://www.jstor.org/stable/1990512},
 author = {Kac, M.},
 journal = {Transactions of the American Mathematical Society},
 number = {1},
 pages = {1--13},
 publisher = {American Mathematical Society},
 title = {On Distributions of Certain Wiener Functionals},
 urldate = {2024-10-09},
 volume = {65},
 year = {1949}
}

@article{Budaev_2005,
  title={A probabilistic approach to wave propagation and scattering},
  author={Budaev, B. V. and Bogy, D. B.},
  journal={Radio science},
  volume={40},
  number={06},
  pages={1--11},
  year={2005},
  publisher={AGU}
}

@INPROCEEDINGS{Kac_1951,
       author = {Kac, M.},
        title = {On Some Connections between Probability Theory and Differential and Integral Equations},
    booktitle = {Second Berkeley Symposium on Mathematical Statistics and Probability},
         year = 1951,
       editor = {Neyman, J.},
        month = jan,
        pages = {189-215},
       adsurl = {https://ui.adsabs.harvard.edu/abs/1951bsms.conf..189K}
}

@INPROCEEDINGS{Kakutani_1944,
       author = {Kakutani, S.},
        title = {Two Dimensional Brownian Motion and Harmonic Function},
    booktitle = {Proceedings of Imperial Academy (Tokyo)},
         year = 1944,
	   volume = 20,
        pages = {706-714},
       adsurl = {https://doi.org/10.3792/pia/1195572706}
}

@article{Feynman_1948,
  title = {Space-Time Approach to Non-Relativistic Quantum Mechanics},
  author = {Feynman, R. P.},
  journal = {Rev. Mod. Phys.},
  volume = {20},
  issue = {2},
  pages = {367--387},
  numpages = {0},
  year = {1948},
  month = {Apr},
  publisher = {American Physical Society},
  doi = {10.1103/RevModPhys.20.367},
  url = {https://link.aps.org/doi/10.1103/RevModPhys.20.367}
}

@article{Kac_1974,
 ISSN = {00357596, 19453795},
 URL = {http://www.jstor.org/stable/44236399},
 author = {Kac, M.},
 journal = {The Rocky Mountain Journal of Mathematics},
 number = {3},
 pages = {497--509},
 publisher = {Rocky Mountain Mathematics Consortium},
 title = {A STOCHASTIC MODEL RELATED TO THE TELEGRAPHER'S EQUATION},
 urldate = {2024-10-09},
 volume = {4},
 year = {1974}
}

@article{Zhang_2019,
title = {Revisiting Kac’s method: A Monte Carlo algorithm for solving the Telegrapher’s equations},
journal = {Mathematics and Computers in Simulation},
volume = {156},
pages = {178-193},
year = {2019},
issn = {0378-4754},
doi = {https://doi.org/10.1016/j.matcom.2018.08.007},
url = {https://www.sciencedirect.com/science/article/pii/S0378475418302052},
author = {Zhang, B. and Yu, W. and Mascagni, M.}
}

@article{Maire_2006,
  TITLE = {{On a Monte Carlo method for neutron transport criticality computations}},
  AUTHOR = {Maire, S. and Talay, D.},
  URL = {https://amu.hal.science/hal-01479840},
  JOURNAL = {{IMA Journal of Numerical Analysis}},
  PUBLISHER = {{Oxford University Press (OUP)}},
  VOLUME = {26},
  NUMBER = {4},
  PAGES = {657-685},
  YEAR = {2006},
  MONTH = Oct,
  HAL_ID = {hal-01479840},
  HAL_VERSION = {v1},
}

@article{Lejay_2010,
  TITLE = {{Simulating diffusions with piecewise constant coefficients using a kinetic approximation}},
  AUTHOR = {Lejay, A. and Maire, S.},
  URL = {https://inria.hal.science/inria-00358003},
  JOURNAL = {{Computer Methods in Applied Mechanics and Engineering}},
  PUBLISHER = {{Elsevier}},
  VOLUME = {199},
  NUMBER = {29-32},
  PAGES = {2014-2023},
  YEAR = {2010},
  MONTH = Jun,
  DOI = {10.1016/j.cma.2010.03.002},
  HAL_ID = {inria-00358003},
  HAL_VERSION = {v4},
}

@article{Bachelier_1900,
     author = {Bachelier, L.},
     title = {Th\'eorie de la sp\'eculation},
     journal = {Annales scientifiques de l'\'Ecole Normale Sup\'erieure},
     pages = {21--86},
     publisher = {Elsevier},
     volume = {3e s{\'e}rie, 17},
     year = {1900},
     doi = {10.24033/asens.476},
     url = {http://www.numdam.org/articles/10.24033/asens.476/}
}

@article{Bachelier_1901,
     author = {Bachelier, L.},
     title = {Th\'eorie math\'ematique du jeu},
     journal = {Annales scientifiques de l'\'Ecole Normale Sup\'erieure},
     pages = {143--209},
     publisher = {Elsevier},
     volume = {3e s{\'e}rie, 18},
     year = {1901},
     doi = {10.24033/asens.493},
     url = {http://www.numdam.org/articles/10.24033/asens.493/}
}

@article{Kac_1947,
author = {Kac, M.},
title = {Random Walk and the Theory of Brownian Motion},
journal = {The American Mathematical Monthly},
volume = {54},
number = {7P1},
pages = {369--391},
year = {1947},
publisher = {Taylor \& Francis},
doi = {10.1080/00029890.1947.11990189},
URL = { 
    
        https://doi.org/10.1080/00029890.1947.11990189
},
eprint = { 
    
        https://doi.org/10.1080/00029890.1947.11990189
}

}

@article{Phillips_1923,
author = {Phillips, H.B. and Wiener, N.},
title = {Nets and the Dirichlet Problem},
journal = {Journal of Mathematics and Physics},
volume = {2},
number = {1-4},
pages = {105-124},
doi = {https://doi.org/10.1002/sapm192321105},
url = {https://onlinelibrary.wiley.com/doi/abs/10.1002/sapm192321105},
eprint = {https://onlinelibrary.wiley.com/doi/pdf/10.1002/sapm192321105},
year = {1923}
}

@article{Lewy_1928,
author = {Lewy, H. and Friedrichs, K. and Courant, R.},
journal = {Mathematische Annalen},
pages = {32-74},
title = {Über die partiellen Differenzengleichungen der mathematischen Physik},
url = {http://eudml.org/doc/159283},
volume = {100},
year = {1928},
}

@article{Haji_1966,
 ISSN = {00361399},
 URL = {http://www.jstor.org/stable/2946271},
 author = {Haji-Sheikh, A. and Sparrow, E.M.},
 journal = {SIAM Journal on Applied Mathematics},
 number = {2},
 pages = {370--389},
 publisher = {Society for Industrial and Applied Mathematics},
 title = {The Floating Random Walk and Its Application to Monte Carlo Solutions of Heat Equations},
 urldate = {2024-10-20},
 volume = {14},
 year = {1966}
}

@inproceedings{Izydorczyk_2019,
  title={McKean Feynman-Kac probabilistic representations of non-linear partial differential equations},
  author={Izydorczyk, L. and Oudjane, N. and Russo, F.},
  booktitle={International Conference on Random Transformations and Invariance in Stochastic Dynamics},
  pages={187--212},
  year={2019},
  organization={Springer}
}

@inbook{Kac_1956,
url = {https://doi.org/10.1525/9780520350694-012},
title = {FOUNDATIONS OF KINETIC THEORY},
booktitle = {Volume 3 Proceedings of the Third Berkeley Symposium on Mathematical Statistics and Probability, Volume III},
author = {M. Kac},
editor = {Jerzy Neyman},
publisher = {University of California Press},
address = {Berkeley},
pages = {173--200},
doi = {doi:10.1525/9780520350694-012},
isbn = {9780520350694},
year = {1956},
lastchecked = {2024-10-17}
}

@article{McKean_1966,
 ISSN = {00278424},
 URL = {http://www.jstor.org/stable/57643},
 author = {McKean, H.P.},
 journal = {Proceedings of the National Academy of Sciences of the United States of America},
 number = {6},
 pages = {1907--1911},
 publisher = {National Academy of Sciences},
 title = {A Class of Markov Processes Associated with Nonlinear Parabolic Equations},
 urldate = {2024-10-09},
 volume = {56},
 year = {1966}
}

@article{McKean_1967,
title = {An exponential formula for solving Boltzmann's equation for a Maxwellian gas},
journal = {Journal of Combinatorial Theory},
volume = {2},
number = {3},
pages = {358-382},
year = {1967},
issn = {0021-9800},
doi = {https://doi.org/10.1016/S0021-9800(67)80035-8},
url = {https://www.sciencedirect.com/science/article/pii/S0021980067800358},
author = {McKean, H.P.}
}

@article{Skorokhod_1964,
author = {Skorokhod, A.V.},
title = {Branching Diffusion Processes},
journal = {Theory of Probability \& Its Applications},
volume = {9},
number = {3},
pages = {445-449},
year = {1964},
doi = {10.1137/1109059},
URL = {https://doi.org/10.1137/1109059},
eprint = {https://doi.org/10.1137/1109059}
}

@misc{Yaacoub_2025,
      title={Nonlinear Drift in Feynman-Kac Theory: Preserving Early Probabilistic Insights}, 
      author={Daniel Yaacoub and Stéphane Blanco and Jean-François Cornet and Jérémi Dauchet and Richard Fournier and Thomas Vourc'h},
      year={2025},
      eprint={2412.08215},
      archivePrefix={arXiv},
      primaryClass={cond-mat.stat-mech},
      url={https://arxiv.org/abs/2412.08215}, 
}

@article{McKean_1975,
author = {McKean, H.P.},
title = {Application of brownian motion to the equation of kolmogorov-petrovskii-piskunov},
journal = {Communications on Pure and Applied Mathematics},
volume = {28},
number = {3},
pages = {323-331},
doi = {https://doi.org/10.1002/cpa.3160280302},
url = {https://onlinelibrary.wiley.com/doi/abs/10.1002/cpa.3160280302},
eprint = {https://onlinelibrary.wiley.com/doi/pdf/10.1002/cpa.3160280302},
year = {1975}
}

@article{Terree_2022,
  title = {Addressing the gas kinetics Boltzmann equation with branching-path statistics},
  author = {Terr\'ee, G. and El Hafi, M. and Blanco, S. and Fournier, R. and Dauchet, J. and Gautrais, J.},
  journal = {Phys. Rev. E},
  volume = {105},
  issue = {2},
  pages = {025305},
  numpages = {20},
  year = {2022},
  month = {Feb},
  publisher = {American Physical Society},
  doi = {10.1103/PhysRevE.105.025305},
  url = {https://link.aps.org/doi/10.1103/PhysRevE.105.025305}
}

@book{Ermakov_1989,
title = "Random Processes for Classical Equations of Mathematical Physics
",
author = "Ermakov, S.M. and Nekrutkin, V.V. and Sipin, A.N.",
year = "1989",
isbn = "978-0-7923-0036-6",
issn = "0169-6378 ",
doi ="https://doi.org/10.1007/978-94-009-2243-3",
publisher = "Springer Dordrecht",
}

@article{Dimov_2000,
title = {Monte Carlo algorithm for solving integral equations with polynomial non-linearity. Parallel implementation},
journal = {Pliska Stud. Math. Bulgar.},
volume = {13},
pages = {117-132},
year = {2000},
author = {Dimov, I.T. and Gurov, T.V.}
}

@Inbook{Kolmogorov_1947,
author={Kolmogorov, A. N. and Dmitriev, N. A.},
editor={Shiryayev, A. N.},
title={Branching Random Processes},
bookTitle={Selected Works of A. N. Kolmogorov: Volume II Probability Theory and Mathematical Statistics (1992)},
year={1947},
publisher={Springer Netherlands},
address={Dordrecht},
pages={309--314},
isbn={978-94-011-2260-3},
doi={10.1007/978-94-011-2260-3_32},
url={https://doi.org/10.1007/978-94-011-2260-3_32}
}

@article{Labordere_2019,
author = {Henry-Labord{\`e}re, P. and Oudjane, N. and Tan, X. and Touzi, N. and Warin, X.},
title = {{Branching diffusion representation of semilinear PDEs and Monte Carlo approximation}},
volume = {55},
journal = {Annales de l'Institut Henri Poincaré, Probabilités et Statistiques},
number = {1},
publisher = {Institut Henri Poincaré},
pages = {184 -- 210},
year = {2019},
doi = {10.1214/17-AIHP880},
URL = {https://doi.org/10.1214/17-AIHP880}
}

@article{Busnello_1999,
 ISSN = {00911798, 2168894X},
 URL = {http://www.jstor.org/stable/2652842},
 author = {Busnello, B.},
 journal = {The Annals of Probability},
 number = {4},
 pages = {1750--1780},
 publisher = {Institute of Mathematical Statistics},
 title = {A Probabilistic Approach to the Two-Dimensional Navier-Stokes Equations},
 urldate = {2024-10-14},
 volume = {27},
 year = {1999}
}

@misc{Nguwi_2023,
	author = {Nguwi, J.Y. and Penent, G. and Privault, N.},
	title = {A fully nonlinear Feynman-Kac formula with derivatives of arbitrary orders},
	year = {2023},
	journal = {Journal of Evolution Equations},
	doi = {10.1007/s00028-023-00873-3},	
	organization ={Ministry of Education (MOE)}
}

@article{Fournie_1999, title={Applications of Malliavin calculus to Monte Carlo methods in finance}, author={Fourni{\'e}, E. and Lasry, J.M. and Lebuchoux, J. and Lions, P.-L. and Touzi, N.}, journal={Finance and Stochastics}, year={1999}, volume={3}, pages={391-412}, url={https://api.semanticscholar.org/CorpusID:6683178} }

@article{Bhattacharya_2003,
title = "Majorizing kernels and stochastic cascades with applications to incompressible Navier-Stokes equations",
author = "Bhattacharya, {Rabi N.} and Larry Chen and Scott Dobson and Guenther, {Ronald B.} and Chris Orum and Mina Ossiander and Enrique Thomann and Waymire, {Edward C.}",
year = "2003",
month = dec,
doi = "10.1090/S0002-9947-03-03246-X",
volume = "355",
pages = "5003--5040",
journal = "Transactions of the American Mathematical Society",
issn = "0002-9947",
publisher = "American Mathematical Society",
number = "12",
}

@article{Fisher_1937,
author = {Fisher, R. A.},
title = {The wave of advance of advantageous genes},
journal = {Annals of Eugenics},
volume = {7},
number = {4},
pages = {355-369},
doi = {https://doi.org/10.1111/j.1469-1809.1937.tb02153.x},
url = {https://onlinelibrary.wiley.com/doi/abs/10.1111/j.1469-1809.1937.tb02153.x},
eprint = {https://onlinelibrary.wiley.com/doi/pdf/10.1111/j.1469-1809.1937.tb02153.x},
year = {1937}
}

@article{Ossiander_2005,
author = {Ossiander, M.},
title = {A probabilistic
representation of solutions of the incompressible Navier-Stokes equations in R3},
volume = {133},
journal = {Probab. Theory Relat. Fields},
pages = {267–298},
year = {2005},
doi = {10.1007/s00440-004-0418-z},
}

@article{Lejan_1997,
author = {Le Jan, Y. and Sznitman, A.},
title = {Stochastic cascades and 3-dimensional Navier–Stokes equations},
volume = {109},
journal = {Probab. Theory Relat. Fields},
pages = {343–366},
year = {1997},
doi = {10.1007/s004400050135},
URL = {https://doi.org/10.1007/s004400050135}
}

@article{Lejay_2020,
title = {A forward-backward probabilistic algorithm for the incompressible Navier-Stokes equations},
journal = {Journal of Computational Physics},
volume = {420},
pages = {109689},
year = {2020},
issn = {0021-9991},
doi = {https://doi.org/10.1016/j.jcp.2020.109689},
url = {https://www.sciencedirect.com/science/article/pii/S0021999120304630},
author = {Lejay, A. and Mardones González, H.}
}

@article{Milstein_2012,
author = {Milstein, G. N. and Tretyakov, M. V.},
title = {Solving the Dirichlet problem for Navier–Stokes equations by probabilistic approach},
year = {2012},
issue_date = {Mar 2012},
publisher = {BIT Computer Science and Numerical Mathematics},
address = {USA},
volume = {52},
number = {1},
issn = {0006-3835},
url = {https://doi.org/10.1007/s10543-011-0347-z},
doi = {10.1007/s10543-011-0347-z},
journal = {BIT},
month = mar,
pages = {141–153},
numpages = {13}
}

@article{Calderoni_1983,
author = {Calderoni, P. and Pulvirenti, M.},
journal = {Annales de l'I.H.P. Physique théorique},
number = {1},
pages = {85-97},
publisher = {Gauthier-Villars},
title = {Propagation of chaos for Burgers' equation},
url = {http://eudml.org/doc/76210},
volume = {39},
year = {1983},
}

@article{Rioux_2022,
  title={A monte carlo method for fluid simulation},
  author={Rioux-Lavoie, D. and Sugimoto, R. and {\"O}zdemir, T. and Shimada, N.H. and Batty, C. and Nowrouzezahrai, D. and Hachisuka, T.},
  journal={ACM Transactions on Graphics (TOG)},
  volume={41},
  number={6},
  pages={1--16},
  year={2022},
  publisher={ACM New York, NY, USA}
}

@inproceedings{Sugimoto_2024,
author = {Sugimoto, R. and Batty, C. and Hachisuka, T.},
title = {Velocity-Based Monte Carlo Fluids},
year = {2024},
isbn = {9798400705250},
publisher = {Association for Computing Machinery},
address = {New York, NY, USA},
url = {https://doi.org/10.1145/3641519.3657405},
doi = {10.1145/3641519.3657405},
booktitle = {ACM SIGGRAPH 2024 Conference Papers},
articleno = {8},
numpages = {11},
location = {Denver, CO, USA},
series = {SIGGRAPH '24}
}

@article{Bienayme_1845,
author="Bienaym\'e, I.J.",
year="1845",
title="De la loi de multiplication et de la durée des familles",
pages="131--132",
issue="589",
journal="L'institut"
}

@article{Galton_1875,
 ISSN = {09595295},
 URL = {http://www.jstor.org/stable/2841222},
 author = {Watson, F.W. and Galton, F.},
 journal = {The Journal of the Anthropological Institute of Great Britain and Ireland},
 number = {},
 pages = {138--144},
 publisher = {[Royal Anthropological Institute of Great Britain and Ireland, Wiley]},
 title = {On the Probability of the Extinction of Families},
 urldate = {2024-07-03},
 volume = {4},
 year = {1875}
}

@article{Jirina_1958,
author = {Jiřina, M.},
doi = {10.21136/CMJ.1958.100304},
issn = {0011-4642},
journal = {Czechoslovak Mathematical Journal},
number = {2},
pages = {292--313},
title = {{Stochastic branching processes with continuous state space}},
url = {https://www.canva.com/policies/terms-of-use/ https://dml.cz/handle/10338.dmlcz/100304},
volume = {08},
year = {1958}
}

@article{Lamperti_1967,
author = {Lamperti, J.},
doi = {10.1007/BF01844446},
journal = {Zeitschrift f{\"{u}}r Wahrscheinlichkeitstheorie und Verwandte Gebiete},
number = {4},
pages = {271--288},
title = {{The Limit of a Sequence of Branching Processes}},
volume = {7},
year = {1967}
}

@article{KPP_1937,
  author = {Kolmogorov, A. and Petrovskii, I. and Piscunov, N.},
  journal = {Byul. Moskovskogo Gos. Univ.},
  number = 6,
  pages = {1--25},
  title = {A study of the equation of diffusion with increase in the quantity of matter, and its application to a biological problem},
  url = {http://books.google.com/books?id=ikN59GkYJKIC&lpg=PP1&dq=A.N.%20Kolmogorov%3A%20Selected%20Works&client=firefox-a&pg=PA242#v=onepage&q=&f=false},
  volume = 1,
  year = 1937
}

@article{Kakutani_1945,
author = {Kakutani, S.},
title = {{Markoff process and the Dirichlet problem}},
volume = {21},
journal = {Proceedings of the Japan Academy},
number = {4},
publisher = {The Japan Academy},
pages = {227 -- 233},
year = {1945},
doi = {10.3792/pja/1195572467},
URL = {https://doi.org/10.3792/pja/1195572467}
}

@article{Warren_2012,
  title = {Malliavin Weight Sampling for Computing Sensitivity Coefficients in Brownian Dynamics Simulations},
  author = {Warren, P.B. and Allen, R.J.},
  journal = {Phys. Rev. Lett.},
  volume = {109},
  issue = {25},
  pages = {250601},
  numpages = {5},
  year = {2012},
  month = {Dec},
  publisher = {American Physical Society},
  doi = {10.1103/PhysRevLett.109.250601},
  url = {https://link.aps.org/doi/10.1103/PhysRevLett.109.250601}
}
\appendix

\end{document}